\documentclass[sigconf,nonacm]{acmart}

\AtBeginDocument{%
  \providecommand\BibTeX{{%
    \normalfont B\kern-0.5em{\scshape i\kern-0.25em b}\kern-0.8em\TeX}}}

\setcopyright{none}

\usepackage[utf8]{inputenc}
\usepackage{booktabs}
\usepackage{multirow}
\usepackage{enumitem}
\usepackage{csquotes}
\usepackage{natbib}

\renewcommand\footnotetextcopyrightpermission[1]{}
\acmConference[CHIWORK '26 Workshop]{%
  CHIWORK '26 Workshop: Interrogating GenAI Augmentation for CHIworkers}{%
  June 22, 2026}{Linz, Austria}
\acmYear{2026}

\begin{document}

\title[Visible Adoption, Untracked Contribution]{Visible Adoption, Untracked Contribution:\\
GitHub Evidence of the Accountability Gap Across\\
Three Cohorts of an HCI Prototyping Course}

\author{Maria Teresa Parreira}
\orcid{0000-0001-6191-3127}
\email{mb2554@cornell.edu}
\affiliation{%
  \department{Information Science}
  \institution{Cornell University}
  \city{New York}
  \state{NY}
  \country{USA}
}

\author{Pranav Prabhat Sinha}
\orcid{0009-0002-8538-8317}
\email{pps46@cornell.edu}
\affiliation{%
  \department{Collaborating Researcher}
  \institution{Cornell Tech}
  \city{New York}
  \state{NY}
  \country{USA}
}

\author{Hauke Sandhaus}
\email{hgs52@cornell.edu}
\orcid{0000-0002-4169-0197}
\affiliation{%
  \department{Information Science}
  \institution{Cornell Tech}
  \city{New York}
  \state{NY}
  \country{USA}
}

\author{Wendy Ju}
\orcid{0000-0002-3119-611X}
\email{wendyju@cornell.edu}
\affiliation{%
  \department{Jacobs Institute}
  \institution{Cornell Tech}
  \city{New York}
  \state{NY}
  \country{USA}
}

\renewcommand{\shortauthors}{Parreira, Sinha, Sandhaus, and Ju}

\begin{abstract}

This paper presents a longitudinal, observational case study of how student GenAI adoption shifted across three cohorts (Fall 2022, 2023, and 2025) of the same graduate-level HCI prototyping course, using computational analysis of 203 GitHub repositories with student activity and 23,065 student commits. Building on a prior qualitative study of the 2023 cohort, we distinguish two levels of AI accountability trace: \emph{disclosure} (naming that an AI tool was used) and \emph{attribution} (crediting a specific artifact or task to an AI tool). We find that tool disclosure grew from 0\% to 66\% of repositories across the three cohorts, while explicit contribution attribution remains a minority practice, and the gap between the two reveals where accountability is missing even among students who disclose.
By 2025, AI is infrastructure embedded in course templates and student-built devices: students increasingly name the tools they used, but rarely specify what those tools contributed. We argue that disclosure-based frameworks are insufficient for the vibe-coding era. The failure is not that students conceal AI use; it is that a norm built for episodic, identifiable acts cannot capture continuous, ambient co-creation. We offer this case study as grounding for the workshop's conversation about what genuine co-thinking accountability looks like.
\end{abstract}


\maketitle

\section{Introduction}

The term ``vibe coding,'' coined by Andrej Karpathy, names a shift that HCI educators have observed in their own classrooms: students increasingly do not \emph{use} AI as a discrete, citable tool but work \emph{within} AI-mediated workflows, where the boundary between the student's contribution and the model's is hard to locate and often goes untracked.
The workshop's \cite{sandhaus2026interrogating} framing of an ``accountability gap'' is relevant here because conventional disclosure (listing the tools one used or the code one borrowed) was designed for settings in which AI assistance was episodic and identifiable. That assumption no longer fits the experience of many HCI students in 2025.

We contribute empirical grounding to this conversation. The Interactive Device Design (IDD) course at Cornell Tech (New York, NY, USA) has run three times under conditions that align with three phases of the Generative AI (GenAI) transition: Fall~2022 (\textit{pre-AI}, no major tools available or used widely), Fall~2023 (\textit{early-AI}, tools like ChatGPT newly available and used more commonly), and Fall~2025 (\textit{vibe-AI}, AI-integrated tools embedded in course infrastructure and student practice). Each cohort followed the same class structure, with a forked GitHub repository maintained through six device-design labs and a final project.
Because the assignment, the workflow scaffold, and the accountability expectation (\textit{``cite what you use''}) remained essentially fixed over the three years, the differences we observe across cohorts are less likely to reflect changes in the course itself and more plausibly reflect changes in the surrounding AI environment.

Our prior qualitative study of the 2023 cohort~\cite{sandhaus_student_2024,sandhaus2025idd} found that all interviewed student groups had used GenAI, mostly beyond what they disclosed, and that usage fell into four \emph{engagement patterns} ranging from AI as a benchmark to AI as a delegated executor. That work documented the accountability gap through student self-report. Here, we ask what \emph{behavioral traces} such as commit messages, file-type distributions, documentation structure, and linguistic patterns reveal about how this gap has evolved. We offer it as an empirical anchor for the workshop's question: what would accountability look like if we designed for it?

\section{Related Work}

GenAI has entered HCI education faster than pedagogy can adapt. Studies document student use of ChatGPT for ideation, code generation, and writing, despite initial reluctance from instructors~\cite{lau_ban_2023, kasneci_chatgpt_2023, baidoo-anu_education_2023}. Empirical accounts of classroom AI integration~\cite{maceli_incorporating_2024, gmeiner_evidence-based_2024} reveal that students routinely adapt AI tools to their workflows in ways instructors did not design for, a form of technology appropriation~\cite{dixDesigningAppropriation2007} that makes top-down policy increasingly untenable. Cotton and colleagues document how academic integrity frameworks strain under ChatGPT use~\cite{cotton_chatting_2024}, while Lau et al.~\cite{lau_ban_2023} argue for temporary bans precisely because no adequate accountability frameworks existed. Bans, however, are increasingly beside the point: AI-mediated coding has become a default condition of software work rather than an optional aid, and students entering the field cannot realistically opt out. The question is therefore no longer whether to permit AI use but how to account for it once it is ambient.

GitHub repositories have been used as behavioral proxies in software engineering research, though Kalliamvakou et al.~\cite{kalliamvakou2014promises} detail their significant limitations as ground truth for developer activity. AI-powered coding assistants like GitHub Copilot create attribution challenges even for professional developers~\cite{bird2022taking}, compounding the difficulty of tracing contribution in educational settings.

Our prior work found that what most shapes the resulting artifact is the \emph{manner} of engagement, not the task or the tool~\cite{sandhaus_student_2024, sandhaus2025idd}. If so, disclosing \emph{what} tool was used and \emph{when} cannot by itself support accountability; the \emph{how} is what current disclosure norms miss. We extend that work computationally and across three cohorts.

\section{Study Context and Method}


The IDD course is a master's-level, project-based course in which students prototype functional interactive devices through six two-week labs (ranging from staging interactions to computer vision or voice-based interactions) and a team-based final project. Each student forks the course's central GitHub repository\footnote{\url{https://github.com/IRL-CT/Interactive-Lab-Hub}} and maintains it throughout the semester. Across all three cohorts, the course policy permitted but did not encourage GenAI use, requiring students to disclose any use, treating AI tools as a citable reference, just as they would cite code from GitHub or instructions from a tutorial~\cite{sandhaus2025idd}. In Fall~2025, the course also shipped course-provided structured AI instruction files (GitHub Copilot chat modes and instruction files) inside every student's fork, intended to feed course context to students' AI tools and to prompt a manual record of AI use. Because that record was optional and we cannot confirm uptake, we treat these files as evidence of course-provided AI \emph{infrastructure} rather than a behavioral log. Cohort sizes were 66~repositories (Fall~2022), 55~repositories (Fall~2023), and 82~repositories (Fall~2025).

\subsection{Repository Analysis}

We fetched all forks of the course repository via the GitHub API, identified 212~course forks by creation date (Aug--Dec of each cohort year), and cloned them locally. Using \texttt{git log}, we extracted 181{,}555 raw commits. These were filtered to a final set of 203~repositories and 23{,}065~student commits by (a)~retaining only commits dated Aug~1--Dec~31 of each cohort year, (b)~excluding commits by 9 class instructors and bots.

\subsection{Operationalization of Analysis Dimensions}

Two levels of AI accountability trace are distinguished throughout. \textbf{Disclosure} (F1) captures whether a student named an AI tool in their README. \textbf{Attribution} (F2) captures sentences that explicitly credit a specific artifact or task to an AI tool (e.g., \textit{``the motor code was written with ChatGPT''}), stating what the tool contributed. The gap between them is where accountability is missing even among students who do disclose.

\textbf{AI tool disclosure in READMEs.}
We searched \texttt{README.md} files in student-created directories, excluding course-template folders (\texttt{WendyTA/}, \texttt{.github/}) which may reference AI tools in their own template descriptions, using case-insensitive exact-string matching for 27~named AI tools (e.g., \textit{chatgpt, claude, gemini, copilot, llama, midjourney, stable diffusion, whisper, langchain}) and 17~AI concept terms (e.g., \textit{llm, language model, system prompt, rag}). The full keyword lists are in the study repository.

\textbf{AI-keyword commits.}
The same keyword list was applied to commit messages. Matches were further filtered against a false-positive exclusion list (\textit{await, grading, repair, certain, maintain, obtain, explain, training data}) to avoid substrings triggering spuriously (e.g., ``cert\underline{ai}n'' contains ``ai''). We also inspected commits for machine-authored attribution metadata such as \texttt{Co-Authored-By} trailers emitted by AI coding agents (e.g., Claude Code, GitHub Copilot), but found no such collaborative commits in any cohort: where AI involvement appears at all, it is expressed in prose rather than in structured commit metadata.

\textbf{File type classification.}
Changed files per commit were classified by extension into four buckets: \textit{code} (e.g., .py, .js, .cpp, .ipynb, .html), \textit{doc} (.md, .txt, .pdf, .docx), \textit{asset} (images, audio, video), and \textit{other}. Percentages are weighted averages across all file changes in a repository (not per-commit averages), to avoid bias from commits that change a single large file type.

\textbf{Commit message style and timing.}
A message was coded \textit{auto/generic} if its first line exactly matched any of 13~low-intentionality patterns (e.g., ``add files via upload,'' ``initial commit,'' merge messages, or single tokens such as \textit{fix, update}). 
\textit{Night} commits were defined as those with a local hour $\geq 23$ or $\leq 5$. Weekend commits were those on Saturday or Sunday per git author timezone. 


\textbf{Explicit AI attribution in READMEs.}
Beyond tool-name presence, we reviewed README text in student-created files for sentences explicitly attributing a contribution to an AI tool: first-person statements (``I used Claude to\ldots''), passive attribution (``code generated by ChatGPT''), or direct credits (``images created with Gemini''). Regex patterns against the 27 tool names were used to surface candidates; all matches were human-verified.

\textbf{Code comment density.}
For each active repository, we walked all Python files excluding virtual-environment directories (\texttt{.venv, venv}), library directories (\texttt{site-packages, lib}), and \texttt{dspeech} --- a 224-file speech-recognition library present in the Fall~2023 course template (48 of 55~\textit{early-ai} repositories; absent in \textit{vibe-ai}). Without this exclusion, early\_ai raw Python file counts inflate from $\sim$58 to $\sim$284 per repository. After correction all three cohorts are comparable ($n = 58/47/60$). Files $>$200KB were skipped. Inline comment lines (starting with \texttt{\#}) and docstring lines (within \texttt{"""}/\texttt{'''}) are expressed as percentages of total code lines.


\textbf{LLM vocabulary markers.}
We assembled 17~phrases across five groups (hedging, discourse, LLM-associated vocabulary, AI self-reference, passive constructions). Counts were normalized per 100~words of README text. Note that this detector was calibrated against 2023-era ChatGPT output; by 2025, models were specifically tuned to avoid these markers, making non-detection in \textit{vibe-ai} uninterpretable as evidence of less AI use.

\textbf{Statistical analysis.}
We use the Mann--Whitney U test throughout. For each metric we report all three pairwise comparisons with Bonferroni correction ($\alpha = 0.05/3 = 0.017$) and rank-biserial correlation ($r_b$) as effect size ($|r_b| > 0.3$ medium, $> 0.5$ large). We report per-repository medians rather than means; file counts use averages. Insertion counts have skewness of 2--5 and are additionally inflated in \textit{early-ai} by upstream template sync commits ($\sim$66K insertions, zero classified file changes, in 77\% of repos). Insertion counts are therefore not compared cross-cohort.

All analyses are available at \url{https://github.com/IRL-CT/idd-x-ai}.
\vspace{-10pt}

\subsection{Limitations}
This is an observational case study of a single course at one institution; \emph{causal attribution to AI tools is not possible}. Differences across cohorts reflect a combination of AI adoption, cohort-level variation, and course template evolution. Specific confounds: (1)~\textit{pre-ai} repositories have a median of 6~unique git authors vs.\ 2--3 for later cohorts, reflecting larger collaborative teams; all per-person metrics normalize by this count, but the proxy is imperfect (multiple git identities per student are counted separately). (2)~The 2025 course template added $\sim$15 new MD files per repository (structured AI instruction files and GitHub Copilot chat modes) absent in earlier years; all README analyses use student-created files only.
(3)~AI signal detection relies on explicit tool name mentions; students who use AI without naming it (e.g., via embedded Copilot suggestions) are systematically undercounted, meaning all disclosure rates reported here are lower bounds. (4)~Cohort skill differences are a plausible confound for F4 and F5: \textit{vibe-ai} students may have entered with stronger programming backgrounds independent of AI use, and this effect cannot be isolated. We corrected for confounds where possible and report only findings that survived scrutiny.

\section{Findings}

Table~\ref{tab:findings} summarizes the key metrics across the three cohorts for active repositories ($\geq 10$~commits; $n = 58$, $47$, $60$ for pre, early, vibe respectively). We report five findings.

\begin{table}[t]
\footnotesize
\setlength{\tabcolsep}{4.5pt}
\renewcommand{\arraystretch}{0.95}
\caption{Key metrics per cohort. Bonferroni-corrected $p$-values; $r_b$ = rank-biserial correlation ($|r_b|>0.3$ medium, $>0.5$ large).}
\label{tab:findings}
\begin{tabular}{lrrr}
\toprule
\textbf{Metric} & \textbf{\textit{pre-ai}} & \textbf{\textit{early-ai}} & \textbf{\textit{vibe-ai}} \\
 & \textit{F'22, n=58} & \textit{F'23, n=47} & \textit{F'25, n=60} \\
\midrule
\multicolumn{4}{l}{\textit{F1 — AI adoption}} \\
\% repos naming AI tools in README & 0\% & 36\% & \textbf{66\%} \\
Median distinct AI tools named & 0 & 2 & \textbf{8} \\
\midrule
\multicolumn{4}{l}{\textit{F2 — Explicit contribution attribution}} \\
\% repos with explicit attribution & 0\% & 23\% & \textbf{42\%} \\
\midrule
\multicolumn{4}{l}{\textit{F3 — Commit activity}} \\
Median commits / person & 26 & 31 & \textbf{49} \\
Median \% auto/generic msgs & 47\% & \textbf{64\%} & 38\% \\
Median \% weekend commits & 11\% & 17\% & \textbf{26\%} \\
Median \% night commits & 11\% & \textbf{19\%} & 8\% \\
\midrule
\multicolumn{4}{l}{\textit{F4 — Work content}} \\
Median \% code file changes & 13\% & 23\% & \textbf{44\%} \\
Median \% asset file changes & 13\% & 23\% & 8\% \\
\midrule
\multicolumn{4}{l}{\textit{F5 — Documentation \& code style}} \\
Median inline comment density & 15.6\% & 14.7\% & \textbf{12.8\%}$\downarrow$ \\
Median docstring density & 2.3\% & 1.9\% & \textbf{3.9\%}$\uparrow$ \\
LLM vocab markers / 100w & 0.028 & \textbf{0.049} & 0.052 \\
\bottomrule
\end{tabular}
\end{table}

\textbf{F1: AI adoption grew monotonically and the tool ecosystem diversified.}
The proportion of repositories naming AI tools in their README, the course's designated disclosure surface, grew from 0\% (\textit{pre-ai}) to 36\% (\textit{early-ai}) to 66\% (\textit{vibe-ai}), with all pairwise differences highly significant ($p < 0.001$, $r_b = 0.68$--$0.94$, large effects). We also observe tool diversification: the median repository names 0 $\to$ 2 $\to$ 8 distinct tools, expanding from a ChatGPT-dominant pattern in \textit{early-ai} to a mix of tools in \textit{vibe-ai} that includes GitHub Copilot, Llama, Moondream, LangChain, and Gemini.

\textbf{F2: Naming tools is not the same as attributing contributions.}
READMEs are the accountability surface in this course, as TAs grade by reading them. Beyond tool-name presence (F1), we searched student-authored README sections for sentences explicitly attributing a specific contribution to an AI tool (first-person statements, passive credits, direct labels): 0\% (\textit{pre-ai}), 23\% (\textit{early-ai}), 42\% (\textit{vibe-ai}) of active repositories contained these sentences. Among \textit{vibe-ai}, we find examples like \textit{``I used Claude to help me tinker the code, I gave it the code and asked to help me set it up''} and \textit{``Gemini assisted during the ideation phase, helping refine written content and parts of the code development''}; \textit{early-ai} attributions are sparser and terser (e.g., \textit{``ChatGPT was used to help adjust Tinkerbelle code''}). The gap between F1 (66\% naming tools) and F2 (42\% attributing contributions) is the accountability gap within the designated disclosure surface itself.

\textbf{F3: Per-person commit frequency rose in \textit{vibe-ai}: consistent with, but not proof of, AI-agent involvement.}
Normalizing by unique authors, \textit{vibe-ai} shows an average of 49~repository commits/person vs.\ 31~(\textit{early-ai}) and 26~(\textit{pre-ai}) ($p < 0.001$, $r_b = +0.57$). The direction is clear but its cause is ambiguous: AI coding agents (Cursor, Claude Code, Copilot) may commit after every change under the student's identity, inflating frequency without added human effort, or AI may simply speed up genuine iteration. The commit log cannot distinguish them.

\textbf{F4: File type composition shifted from assets to code.}
The percentage of code file changes grew from a median of 13\% (\textit{pre-ai}) to 44\% (\textit{vibe-ai}) ($p < 0.001$, $r_b = +0.81$, large). Since the lab assignments are largely unchanged (students still build Raspberry Pi devices for comparable tasks), this reflects a change in how students work, with \textit{pre-ai} students documenting largely with photographs of hardware, whereas \textit{vibe-ai} students produced substantially more code for the same tasks.

\textbf{F5: Code structure and writing style diverge in \textit{vibe-ai}.}
Two signals emerge. First, \textit{vibe-ai} code has significantly \emph{fewer} inline comments (median 12.8\%) than \textit{pre-ai} (15.6\%; $p < 0.001$, $r_b = -0.64$) and \textit{early-ai} (14.7\%; $p < 0.001$, $r_b = -0.54$). Simultaneously, \textit{vibe-ai} has significantly \emph{more} docstrings (3.9\%) than \textit{pre-ai} (2.3\%; $p < 0.001$, $r_b = +0.65$) and \textit{early-ai} (1.9\%; $p < 0.001$, $r_b = +0.72$). Together, these suggest a code style shift, with LLMs reliably adding structured docstrings to functions while omitting the inline \texttt{\# do this} comments characteristic of novice coders working from tutorials. LLM-vocabulary markers in READMEs show a significant step-change between \textit{pre-ai} and \textit{early-ai} ($p < 0.001$, $r_b = +0.69$) and then plateau (early vs.\ vibe: $p = 0.20$, ns), likely because by 2025 models were tuned to avoid these markers.

\section{Discussion: What This Means for Accountability}

\noindent\textbf{Adoption is visible; contribution is not.}
Tool naming (66\% of \textit{vibe-ai} repositories) and explicit contribution attribution (42\%) measure different things: the first shows that students acknowledge AI's presence, the second that they credit specific work to it. Yet even attribution is coarse. A phrase like ``edited with AI'' collapses two quite different practices: feeding an entire lab to a model and committing its one-shot output, versus iterating over code across many human-directed passes. The F2 examples span exactly this range, but the README sentence alone rarely lets a reader tell which occurred.
Even within the intended accountability surface (the README), naming tools is more common than attributing contributions; moreover, a bare disclosure such as ``Copilot was used for coding'' says little about how much of the work, or which decisions, remained the student's.
The accountability gap is not primarily about students hiding AI use; it is about a disclosure norm designed for episodic, attributable acts (``\textit{I used Claude for this function}'') struggling to capture continuous, ambient co-creation.

\noindent\textbf{AI has become infrastructure, not a tool.}
By Fall~2025, every student fork shipped with course-provided AI instruction files and pre-configured GitHub Copilot chat modes, meaning AI was embedded in the course template before any student chose to use it. Accountability frameworks that place the burden of disclosing AI contribution on the student cannot easily handle AI supplied as part of the educational scaffold. This widens the gap \emph{structurally}, not just behaviorally: the behavioral gap is a student not disclosing a tool they chose to use, whereas the structural gap is AI present by default, so that even a fully compliant student has nothing to ``disclose.'' Disclosure regimes presume an opt-in act to report; infrastructure that embeds AI removes the act of disclosure while keeping the contribution.

\noindent\textbf{Increased output does not resolve the co-thinking question.}
The F3/F4/F5 cluster presents the same interpretive problem from three angles. More commits per person (F3), more code relative to assets (F4), and longer, more structured documentation (F5) are all consistent with either AI amplifying student capability or AI performing the work while students supervise. An AI agent committing under a student's identity still records human authorship; a pasted ChatGPT docstring still looks professionally structured. Our traces cannot distinguish co-thinking from outsourcing. If intellectual autonomy is the goal, accountability mechanisms need to operate at a different layer, one that probes the reasoning behind design decisions, not just their outputs.

\noindent\textbf{Position: Toward architecture-level accountability.}
We propose that ``did you document your AI use?'' is the wrong question, and that accountability for the vibe-coding era should operate at the \emph{architectural} level: what does the artifact do with AI, and could the student explain why?
The clearest signal in our data was not disclosure but the structure of what students built. Prior work distinguishes four \emph{engagement patterns} (how a student relates to the model, from benchmark to delegated executor) from nine \emph{activity types} (the kind of work AI does: brainstorming, prototyping, coding, hardware integration, reflection)~\cite{sandhaus_student_2024, sandhaus2025idd}, and these map to very different degrees of student agency. A student who wires a language model into a multi-step pipeline (e.g., LangChain) to build a device assistant that reasons over its own documentation has made design decisions a disclosure form cannot capture; a student who pasted a ChatGPT answer into a commit message has not. Both look identical if disclosure of use is the only mechanism for accountability.

\section{Conclusion}

Three cohorts of the same course provide longitudinal evidence of how AI integration and its accountability traces have evolved. AI adoption grew and diversified (F1); explicit contribution attribution grew more slowly and remained a minority practice (F2); commit frequency and code volume increased in ways consistent with AI-agent involvement but not attributable to it (F3, F4); and writing style signals suggest AI assistance that increasingly evades detection (F5). The accountability gap is real and observable, and may widen further as models improve. This case study is offered to the workshop as empirical grounding that disclosure-based frameworks are necessary but insufficient.

\section*{Use of AI Tools}
AI tools (Claude) were used to help write and debug the data-extraction and analysis scripts, to assist exploratory analysis, and to refine the writing. The study design, choice of metrics, statistical decisions, and all findings and claims were determined and human-verified by the authors~\cite{ACM-Publications-Board2023-wp}. We report this in the spirit of the disclosure norm the paper examines.

\begin{acks}
We thank all IDD students across the three cohorts. We also thank the reviewers of the DIS~'25 paper on which this work builds. Data collection was approved as exempt by the Cornell University Human Participant Review Board under IRB0148132.
\end{acks}

\bibliographystyle{abbrv}
\bibliography{GenAI-prototyping-class,manual-references}

\end{document}